\documentclass{iacrtrans}
\usepackage[utf8]{inputenc}
\usepackage{cite}
\usepackage{amsmath,amssymb,amsfonts}
\usepackage{graphicx}
\usepackage{textcomp}
\usepackage{xcolor}
\def\BibTeX{{\rm B\kern-.05em{\sc i\kern-.025em b}\kern-.08em
    T\kern-.1667em\lower.7ex\hbox{E}\kern-.125emX}}
    
\input epsf
\usepackage{listings}
\usepackage{amsmath}
\usepackage{amsfonts,amssymb}
\DeclareSymbolFontAlphabet{\mathbb}{AMSb}

\usepackage{cite} 
\usepackage[ruled,linesnumbered]{algorithm2e}
\usepackage{float}
\usepackage{graphicx}

\SetCommentSty{mycommfont}

\author{Rashmi Agrawal\inst{1}, Lake Bu\inst{2}, Alan Ehret\inst{1}  \and Michel A. Kinsy\inst{1}}
\institute{Adaptive and Secure Computing Systems (ASCS) Laboratory, Boston University \email{rashmi23, ehretaj, mkinsy@bu.edu} \and
          The Charles Stark Draper Laboratory Inc., USA, \email{lbu@draper.com}}
\title{Fast Arithmetic Hardware Library For RLWE-Based Homomorphic Encryption}

\begin{document}

\maketitle


\begin{abstract}
With billions of devices connected over the internet, the rise of sensor-based electronic devices has led to the use of cloud computing as a commodity technology service. These sensor-based devices are often small and limited by power, storage, or compute capabilities; hence, they achieve these capabilities via cloud services. However, this heightens data privacy issues as sensitive data is stored and computed over the cloud, which, at most times is a shared resource. Homomorphic encryption can be used along with cloud services to perform computations on encrypted data, guaranteeing data privacy. While work on improving homomorphic encryption has ensured its practicality, it is still several magnitudes too slow to make it cost effective and feasible. In this work, we propose an open-source, first-of-its-kind, arithmetic hardware library with a focus on accelerating the arithmetic operations involved in Ring Learning with Error (RLWE)-based somewhat homomorphic encryption (SHE). We design and implement a hardware accelerator consisting of submodules like Residue Number System (RNS), Chinese Remainder Theorem (CRT), NTT-based polynomial multiplication, modulo inverse, modulo reduction, and all the other polynomial and scalar operations involved in SHE. For all of these operations, wherever possible, we include a hardware-cost efficient serial and a fast parallel implementation in the library. A modular and parameterized design approach helps in easy customization and also provides flexibility to extend these operations for use in most homomorphic encryption applications that fit well into emerging FPGA-equipped cloud architectures. Using the submodules from the library, we prototype a hardware accelerator on FPGA. The evaluation of this hardware accelerator shows a speed up of approximately $4200\times$ and $2950\times$ to evaluate a homomorphic multiplication and addition respectively when compared to an existing software implementation.  
\end{abstract}

\section{Introduction}
\label{Intro}
As the internet becomes easily accessible, almost all electronic devices collect enormous amounts of private and sensitive data from routine activities. These electronic devices may be as small as wearable electronics \cite{PE2003}, like a smart watch collecting personal health information or a cell phone collecting location information, or may be as large as an IoT-based smart home \cite{BD2011} \cite{KI2016} collecting routine information like room temperature, door status (open or closed), smart meter reading, and other such details. These electronic devices have limited power, storage, and compute capabilities and often need external support to process the collected information. Cloud computing \cite{MN2011} \cite{HC2008} provides a convenient means not only to store the collected information but also to apply various compute functions to this stored data. The processed information can be used easily for various purposes, like machine learning predictions. 

As cloud computing services become readily available and affordable, many industrial sectors have begun to use cloud services instead of setting up their own infrastructure. Sectors like automotive production, education, finance, banking, health care, manufacturing, and many more leverage cloud services for some or all of their storage and computing needs. This, in turn, leads to the storage of a lot of sensitive data in the cloud. Hence, while cloud computing provides the convenience of sharing resources and compute capabilities for individuals and business owners equally, it brings its own challenges in maintaining data privacy \cite{PC2010} \cite{KC2011}. A cloud owner has access to all of the private data pertaining to the clients and can also observe the computations being carried out on this private data. Moreover, cloud services are shared among many clients; therefore, even if a client may assume that the cloud service provider is honest and will ensure that there is no data breach in their environment, the chances of data leakage remain high, due to the shared storage space or compute node on the cloud. 

\begin{figure}[ht]
\begin{center}
	\includegraphics[width=0.75\columnwidth]{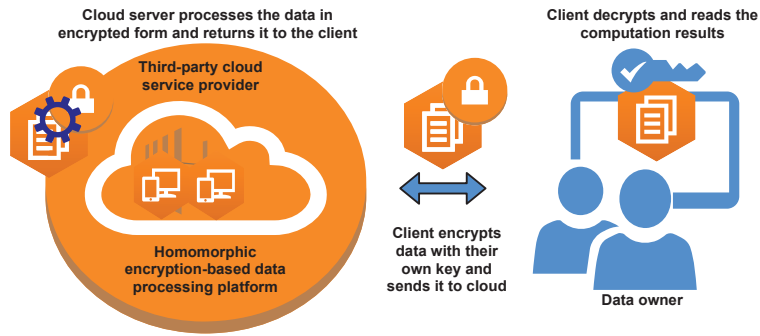} 
\end{center}
\vspace{-0.1in}
	\caption{Third-party cloud service provider with Homomorphic Encryption.} 
	\label{fig:cloud}
	\vspace{-0.05in}
\end{figure}

Homomorphic encryption \cite{GF2009} is a ground-breaking technique to enable secure private cloud storage and computation services. Homomorphic encryption allows evaluating functions on encrypted data to generate an encrypted result. This result, when decrypted, matches the result of the same operations performed on the unencrypted data. Thus, a data owner can encrypt the data and then send it to cloud for processing. The cloud running the homomorphic encryption based services will perform computations on the encrypted data and send the results back to the data owner. The data owner, having access to the private key, performs the decryption and obtains the result. The cloud does not have access to the private key or the plain data, and, hence, the security concerns related to private data processing on the cloud can be mitigated. An illustrative scenario is shown in Figure \ref{fig:cloud}.

The idea of homomorphic encryption was first proposed in 1978 by Rivest et al. \cite{RD1978}. In 2009, Gentry's seminal work \cite{GF2009} provided a framework to make fully homomorphic encryption feasible, and almost a decade's work has now made it practical \cite{NC2011}. While homomorphic encryption has become realistic, it still remains several magnitudes too slow, making it expensive and resource intensive. There are no existing homomorphic encryption schemes with performance levels that would allow large-scale practical usage. Substantial efforts have been put forward to develop full-fledged software libraries for homomorphic encryption. Such libraries include SEAL \cite{CS2017}, Palisade \cite{PL2019}, cuHE \cite{DC2015}, HElib \cite{HS2014}, NFLLib \cite{AN2016}, Lattigo \cite{LT2019}, and HEAAN \cite{KH2018}. All of these libraries are based on the RLWE-based encryption scheme, and they generally implement Brakerski-Gentry-Vaikuntanathan (BGV) \cite{BL2014}, Fan-Vercauteren (FV) \cite{FS2012}, and Cheon-Kim-Kim-Song (CKKS) \cite{CH2017} homomorphic encryption schemes with very similar parameters. 

Although the software implementations are impressive, they are still incapable of gaining the required performance, as they are limited by the underlying hardware. For example, Gentry et al. \cite{GC2012}, in their homomorphic evaluation of an AES circuit, reported approximately $48$ hours of execution time on an Intel Xeon CPU running at $2.0$GHz. Even their parallel SIMD style implementation took around $40$ minutes per block to evaluate $54$ AES blocks. A modern Intel Xeon CPU takes about $20$ns to perform a regular AES encryption block, hence it is evident that homomorphic evaluation of an AES block is about $1.2\times10^{11}$ times slower than a regular evaluation. Similarly, logistic regression, a popular machine learning tool, is often used to make predictions using client's private data in the cloud. A software-based homomorphic logistic regression prediction takes about $1.6$ hour while a regular logistic prediction takes about $95$ns.    

If homomorphic encryption's full potential and power can be unleashed by realizing the required performance levels, it will make cloud computing more reliable via enhanced trust of service providers and their mechanisms for protecting users’ data. Hence, there is a need to accelerate the homomorphic encryption operation directly on the hardware to achieve maximum throughput with a low latency. With this in mind, we propose an arithmetic hardware library that includes the major arithmetic operations involved in homomorphic encryption. A hardware accelerator designed using the modules from this library can reduce the computational time for HE operations. To lower the power usage and improve performance, new cloud architectures integrate FPGAs to offload and accelerate compute tasks such as deep learning, encryption, and video conversion. The FPGA-based design and optimization approach introduced in this work fits into this class of FPGA-equipped cloud architectures. 

The key contributions of the work are as follows:
\begin{itemize}
    \item A fast and hardware-cost efficient hardware arithmetic library to individually accelerate all operations within homomorphic encryption. A speedup of $4200\times$ and $2950\times$ is observed to evaluate homomorphic multiplication and addition respectively. 
    \item An open-source, FPGA-board agnostic, parameterized design implementation of the modules to provide flexibility to adjust parameters so as to meet the desired security levels, hardware cost and multiplication depth.
    \item A modular and hierarchical implementation of a hardware accelerator using the modules of the proposed arithmetic library to demonstrate the speedup achievable in hardware. 
\end{itemize}

The rest of the paper is organized as follows. In Section \ref{HE}, we briefly present the underlying scheme and discuss the required arithmetic operations. Section \ref{FPGA} introduces these arithmetic operations and their efficient implementation. In Section \ref{Perf}, we evaluate the associated hardware cost and latency and then conclude the paper in Section \ref{con} along with future work. 
\section{Homomorphic Encryption}
\label{HE}
To present the hardware library, we first start by introducing the underlying RLWE-based homomorphic encryption scheme. For this purpose, we chose the Fan-Vercauteren (FV) \cite{FS2012} scheme, as it has more controlled noise growth while performing homomorphic operations when compared to approaches like the BGV (Brakerski-Gentry-Vaikuntanathan) scheme~\cite{BL2014}. Moreover, Costache et al. \cite{CE20} presented results showing FV scheme outperforming BGV for large plaintext moduli.

\subsection{FV Scheme}
The FV scheme operates in the ring $R = \mathbb Z_Q[x]/ \left< f(x) \right>$, with $f(x) = \phi_d(x)$ the $d^{th}$ cyclotomic polynomial. The plaintext $m$ is chosen in the ring $R_t$ for some small $t$ and a ciphertext consists of only one element in the ring $R_Q$ for a large integer $Q$. The security of the scheme is governed by the degree of this polynomial $f(x)$ and the size of $Q$.  

The secret key, $s_k$, is sampled from the ring $R_2$ or a Gaussian distribution, $\chi_k$. The public key, $a$, is sampled from the ring $R_Q$ and the error vector, $e$, is sampled from a second Gaussian distribution, $\chi_{err}$. The other public key, $b$, is computed as follows:
\begin{equation}
    b = [-(a \cdot s + e)]_{R_Q}
    \label{equ:KeyGen}
\end{equation}

Encryption of the plaintext message yields a pair of ciphertexts as follows:
\begin{equation}
    ct = ([(b \cdot r_0 + r_2 + t \cdot m)]_{R_Q}, [(a \cdot r_0 + r_1)]_{R_Q})
    \label{equ:Enc}
\end{equation}

The homomorphic addition operation adds two such pairs of ciphertexts:
\begin{equation}
  (c_0, c_1) = ([ct_1[0] + ct_2[0]]_{R_Q}, [ct_1[1] + ct_2[1]]_{R_Q})
  \label{equ:Add}
\end{equation}

After the addition operation, decryption is done as:
\begin{equation}
    m_1 + m_2 = \left[ \bigg \lfloor \frac{[c_0 + c_1 \cdot s_k]}{t} \bigg \rfloor \right]_{R_Q}
    \label{equ:AddDec}
\end{equation}

The homomorphic multiplication operation multiplies two such pairs of ciphertexts using the following equations:
\begin{align}
\begin{split}
    c_0 = [ct_1[0] \cdot ct_2[0]]_{R_Q} \\
    c_1 = [ct_1[0] \cdot ct_2[1] + ct_1[1] \cdot ct_2[0]]_{R_Q} \\
    c_2 = [ct_1[1] \cdot ct_2[1]]_{R_Q} 
    \label{equ:Mul}
\end{split}
\end{align}

Decryption, after the multiplication operation, using the secret key is computed as:
\begin{equation}
    m_1 \cdot m_2 = \left[ \bigg \lfloor \frac{[c_0 + c_1 \cdot s_k + c_2 \cdot s^2_k]}{t} \bigg \rfloor \right]_{R_Q}
    \label{equ:MulDec}
\end{equation}

Since after the multiplication operation a degree $2$ ciphertext is obtained, to continue further multiplication operations this degree $2$ ciphertext needs to be reduced to a degree $1$ ciphertext. In the FV scheme, this is achieved by performing a relinearisation operation. The scheme facilitates two different approaches for performing relinearisation. Relinearisation version $1$ operation consists of generating the relinearisation key, decomposing $c_2$ to limit the noise explosion, and then conversion to a degree $1$ ciphertext by using generated relinearisation keys and the decomposed $c_2$. The relinearisation keys are generated as follows:
\begin{equation}
    rlk = [([-(a_i \cdot s + e_i) + T^i \cdot s^2]_{R_Q},a_i) : i \in [0..\ell]]
    \label{equ:Relin1KeyGen}
\end{equation}

Here, $T$ is independent of $t$ and $\ell = \lfloor log_T(Q) \rfloor$. Decomposition of $c_2$ involves rewriting $c_2$ in base $T$ and can be computed using the following equation:
\begin{equation}
    c_2 = \sum_{i=0}^{\ell} T^i \cdot c_{2}^{(i)}
    \label{equ:Decomp}
\end{equation}

Next the relinearisation operation can be performed as follows:
\begin{align}
\begin{split}
    c_{0}^{'} = [c_0 + \sum_{i=0}^{\ell} rlk[i][0] \cdot c_{2}^{(i)}]_{R_Q} \\
    c_{1}^{'} = [c_1 + \sum_{i=0}^{\ell} rlk[i][1] \cdot c_{2}^{(i)}]_{R_Q}
    \label{equ:Relin1}
\end{split}
\end{align}

Since, we have obtained a degree $1$ ciphertext after the relinearisation operation, the decryption can be performed without using the $s_{k}^{2}$ term, as in equation \ref{equ:MulDec}. Therefore, the decryption operation simplifies to the equation:
\begin{equation}
    m_1 \cdot m_2 = c_0^{'} + c_1^{'} \cdot s_k
    \label{equ:RelinMulDec}
\end{equation}

Note that the choice of $T$ will determine the size of relinearisation keys and the noise growth during the relinearisation operation. The larger the value of $T$, the smaller the relinearisation keys will be, but the noise introduced by relinearisation will be higher. And the smaller the value of $T$, the larger the relinearisation keys will be, with smaller noise introduction. So the value of $T$ must be picked in a balanced way. 

Relinearisation version $2$ is a modified form of modulus switching and hence requires choosing a second modulus $p$ such that $p \geq Q^3$ for small enough error samples. Now, the relinearisation keys can be generated as follows:
\begin{equation}
    rlk = ([-(a \cdot s + e) + p \cdot s^2]_{R_{p \cdot Q}}, a)
    \label{equ:Relin2KeyGen}
\end{equation}
Here, $a \in R_{p \cdot Q}$ and $e \xleftarrow{} \chi'_{err}$. We can perform the relinearisation using the following computation:
\begin{align}
\begin{split}
    c_{0}^{'} = c_0 + \left[ \bigg \lfloor \frac{c_{2} \cdot rlk[0]}{p} \bigg \rceil \right] _{R_Q} \\
    c_{1}^{'} = c_1 + \left[ \bigg \lfloor \frac{c_{2} \cdot rlk[1]}{p} \bigg \rceil \right]_{R_Q}
    \label{equ:Relin2}
\end{split}
\end{align}
Once the $c_2$ component is removed, we can perform decryption using equation \ref{equ:RelinMulDec}. 

\subsection{Required Operations}
Our proposed arithmetic library includes highly optimized hardware-based implementations of Residue Number System (RNS), Chinese Remainder Theorem (CRT), modulo inverse, fast polynomial multiplication using Number Theoretic Transform(NTT), polynomial addition, modulo reduction, Gaussian noise sampler and relinearisation operations. While implementing these operations for our arithmetic library, the design choices are highly motivated by the parameter selection. This is because an RLWE-based encryption scheme requires adding a small noise vector to obfuscate the plaintext message as shown previously in equations \ref{equ:KeyGen} and \ref{equ:Enc}. While performing homomorphic addition and multiplication, the noise present in the ciphertexts gets doubled and squared respectively. Due to this noise growth, the ring $R_Q$, along with the degree of the polynomial needs to be large, so as to compute a circuit of certain depth and still enable successful decryption of the result. Hence, the parameter $Q$ with the degree of the polynomial $f(x)$ needs to be large. The operations on this large parameter set not only increase the hardware cost but also slow down the homomorphic encryption.    

\begin{figure}[ht]
\begin{center}
	\includegraphics[width=0.95\columnwidth]{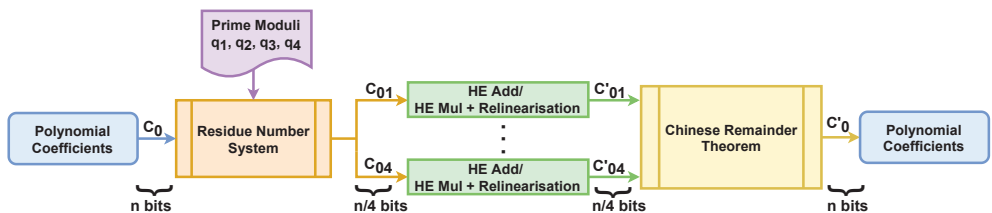} 
\end{center}
\vspace{-0.15in}
	\caption{Illustrative sequence of operations.} 
	\label{fig:RO}
\end{figure}

\begin{figure}[H]
\begin{center}
	\includegraphics[width=1.0\textwidth]{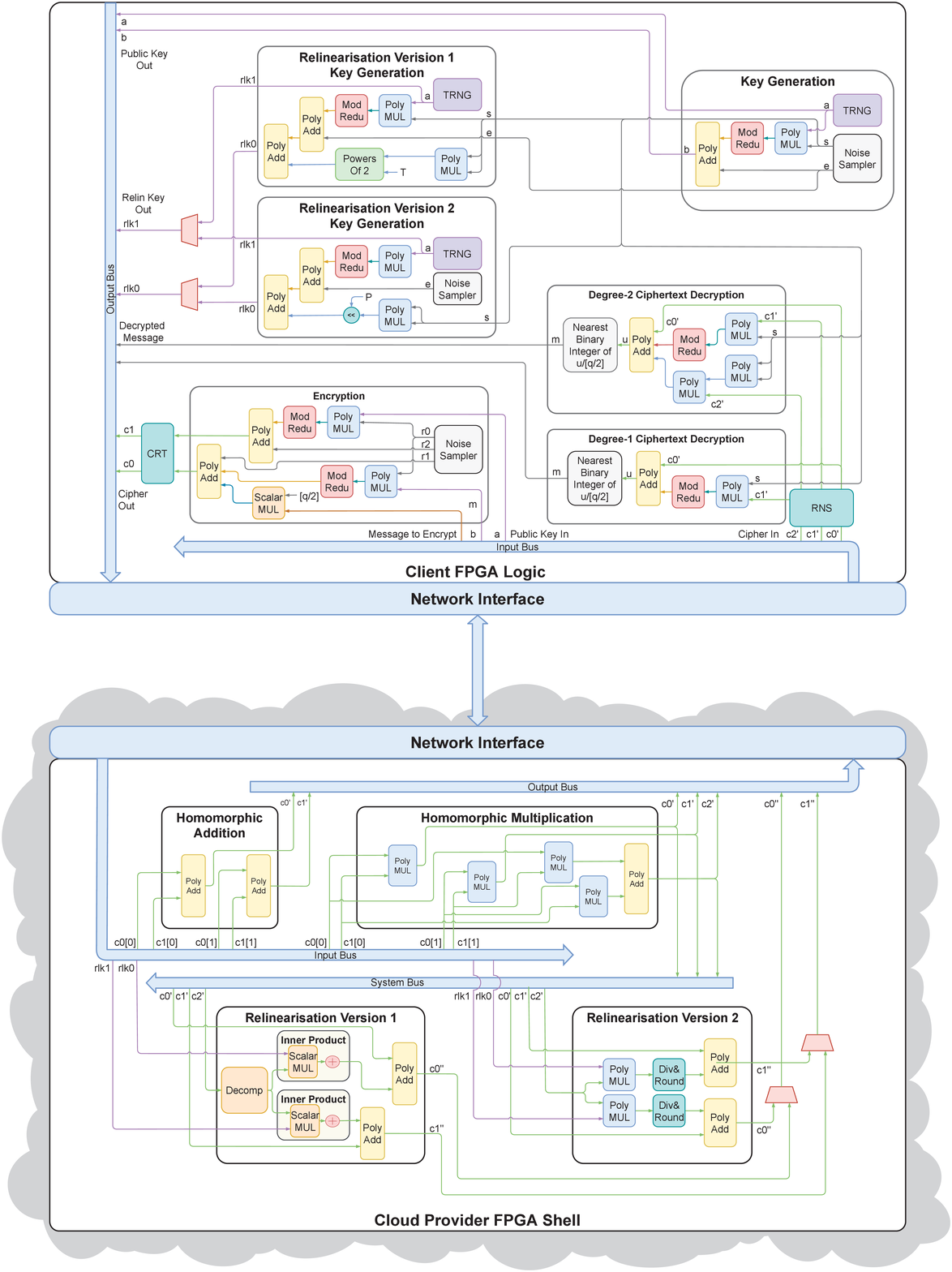}
\end{center}
\vspace{-0.1in}
	\caption{Core building blocks of RLWE-based Somewhat Homomorphic Encryption.} 
	\label{fig:SHE}
\end{figure}

We use the concepts of modular arithmetic to speed up HE computations. The underlying FV scheme does not restrict $Q$ to a prime number; instead, $Q$ can be the product of small primes. When working modulo a product of numbers, say $Q = q_1 \times q_2 \times \cdots \times q_k$, Residue Number System (RNS) helps reduce the coefficients in each of the $R_{q_i}$ and Chinese Remainder Theorem (CRT) lets us work in each modulus $q_i$ separately. Since the computational cost is proportional to the size of operands, this is faster than working in the full modulus $Q$. Moreover, breaking down the coefficients into smaller integers using RNS also limits the noise expansion. Figure \ref{fig:RO} illustrates the sequence of operations that will be required in HE using the FV scheme, for $Q = q_1 \times q_2 \times q_3 \times q_4$.

In our implementation of the arithmetic library, we consider $Q$ to be $1200$ bits and degree of the polynomial, $n$ as $1024$ with a $128$-bit security level. With these values of $Q$ and $n$, we can evaluate a binary circuit of depth $56$ using somewhat homomorphic encryption. The RNS module will take a $1200$-bit wide integer coefficient as an input, perform $x_i = x \ mod \ q_i$, and thus break $x$ into $40$ small integers of $30$ bits each. This enables us to set up $40$ pipelines to perform $40$ operations in parallel, providing the required performance boost. Once the homomorphic add or multiplication operation is done on these small integers, CRT can combine them to map back to the original $1200$-bit width. Note that the bit width selection for the small integers is a design decision one can make based on available resources. Our implementations take $q_i$ as a parameter, which facilitates different bit width selections. 

Figure \ref{fig:SHE} shows all the core building blocks required to perform somewhat homomorphic encryption (SHE) addition and multiplication operations using FV scheme. The client-side building blocks include key generation, relinearisation (versions $1$ and $2$) key generation, encryption, decryption (for both degree-1 and degree-2 ciphertexts), residue number system (RNS) along with modular reduction, and Chinese remainder theorem (CRT) along with modulo inversion. The cloud provider has blocks to perform homomorphic addition, homomorphic multiplication, and relinearisation (versions $1$ and $2$). While RNS and CRT are standalone modules, the rest of the main blocks share the following submodules: polynomial multiplication, polynomial addition, modular reduction, scalar multiplication, scalar division to nearest binary integer, noise sampler, and true random number generator. Certain operations like decompositions, powers-of-$2$ computations, divide and round operations are specifically required for the purpose of relinearisation, which we will discuss in detail in Section \ref{relin}.

\section{Arithmetic Hardware Library For HE}
\label{FPGA}
We will start with discussing the standalone modules first (i.e., RNS and CRT) along with the supplemental operations to these modules. Then, we will describe other arithmetic operations shared between all the core building blocks, along with their design implementations. All the operations are customized for hardware-based implementation, and we include both hardware-cost efficient serial and fast parallel implementations in the library. It is worth noting that, in all the algorithms, $Q$ will denote the large integer and $q_i$ or $q$ will denote a prime factor of $Q$.

\subsection{Residue Number System}
A residual number system, RNS \cite{GR1959} \cite{SR1986} \cite{SR1967} is a mathematical way of representing an integer by its value modulo a set of k integers ${q_1,q_2,q_3,\ldots ,q_k}$, called the moduli, which generally should be pairwise coprime. An integer, $x$, can be represented in the residue number system by a set of its remainders ${x_1,x_2,x_3,\ldots ,x_k}$ under Euclidean division by the respective moduli. That is, $x_i = x \pmod q_i$ and $0\leq x_{i}<q_{i}$ for every $i$.

\begin{figure}[ht]
\begin{center}
	\includegraphics[width=0.85\columnwidth]{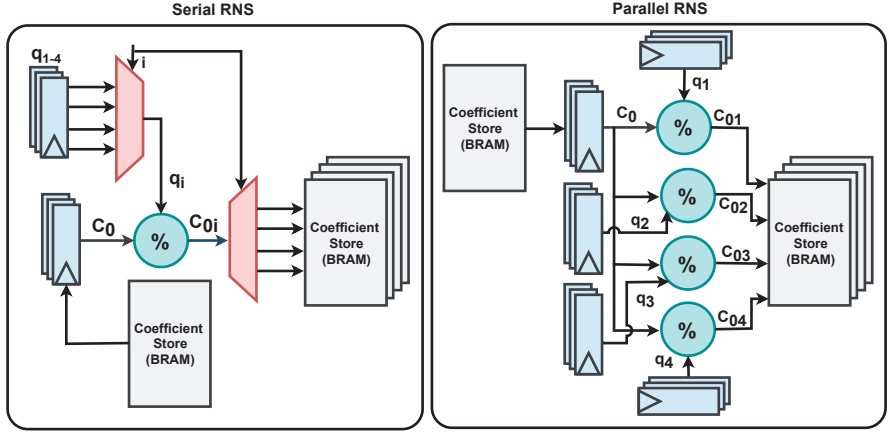} 
\end{center}
	\caption{Serial and parallel implementation of RNS.} 
	\label{fig:RNS}
\end{figure}

The serial implementation of RNS is shown in Figure \ref{fig:RNS}. Each $1200$-bit coefficient is modulo reduced by a $q_i$ and stored in the respective BRAM. For $k$ moduli, it takes $k$ cycles to perform all the computations. When modulo reductions are performed in parallel, all the computations can be performed in a single clock cycle instead. The parallel implementation is shown in Figure \ref{fig:RNS}. Since the $mod$ operation is the key operation in RNS, we next optimize the modulo reduction operation. This will allow us to reduce the hardware cost substantially.

\subsection{Modular reduction}
Modular reduction is not only at the core of many asymmetric cryptosystems, it is the most performed operation in encryption schemes based on R-LWE. This is because, in RLWE, all the operations are required to be performed over large finite rings. The function of the modular reduction operation is to compute the remainder of an integer division. Mathematically, it is written as $r = a \pmod q$.

While the modular reduction operation sounds relatively simple, the division of two large integers is very costly. Moreover, the moduli in RLWE-based schemes are prime numbers and not power-of-2 numbers, which makes the operation non-trivial. Therefore, the hardware implementation of the modulo operation is quite expensive. For example, the use of the inbuilt modulo operator, $\%$ in Verilog for $30$-bit operands utilizes about $800$ LUTs, and when there are many such modular operations involved, the hardware cost quickly adds up. Hence, optimization of modulo operation can lead to significant hardware cost reductions. 

One well-known modular reduction optimization algorithm is Barrett reduction\cite{BP1986}. It is preferred over Montgomery reduction \cite{MM1985}, as it operates on the given integer number directly, while Montgomery reduction requires numbers to be converted into and out of Montgomery form, which is expensive in itself. We will discuss the Barrett reduction next, and then later, we propose some modifications to the existing Barrett reduction algorithm to reduce the hardware cost further.              

\subsubsection{Barrett reduction}
The Barrett reduction algorithm was introduced by P. D. Barrett \cite{BP1986} to optimize the modular reduction operation by replacing divisions with multiplications, so as to avoid the slowness of long divisions. The key idea behind the Barrett reduction is to precompute a factor using division for a given prime modulus, $q$, and thereafter, the computations only involve multiplications, subtractions and shift operations. These operations are faster than the division operation. The Barrett reduction algorithm steps are shown in Algorithm \textbf{1} and works as described below.

\begin{figure}[ht]
\begin{center}
	\includegraphics[width=0.99\columnwidth]{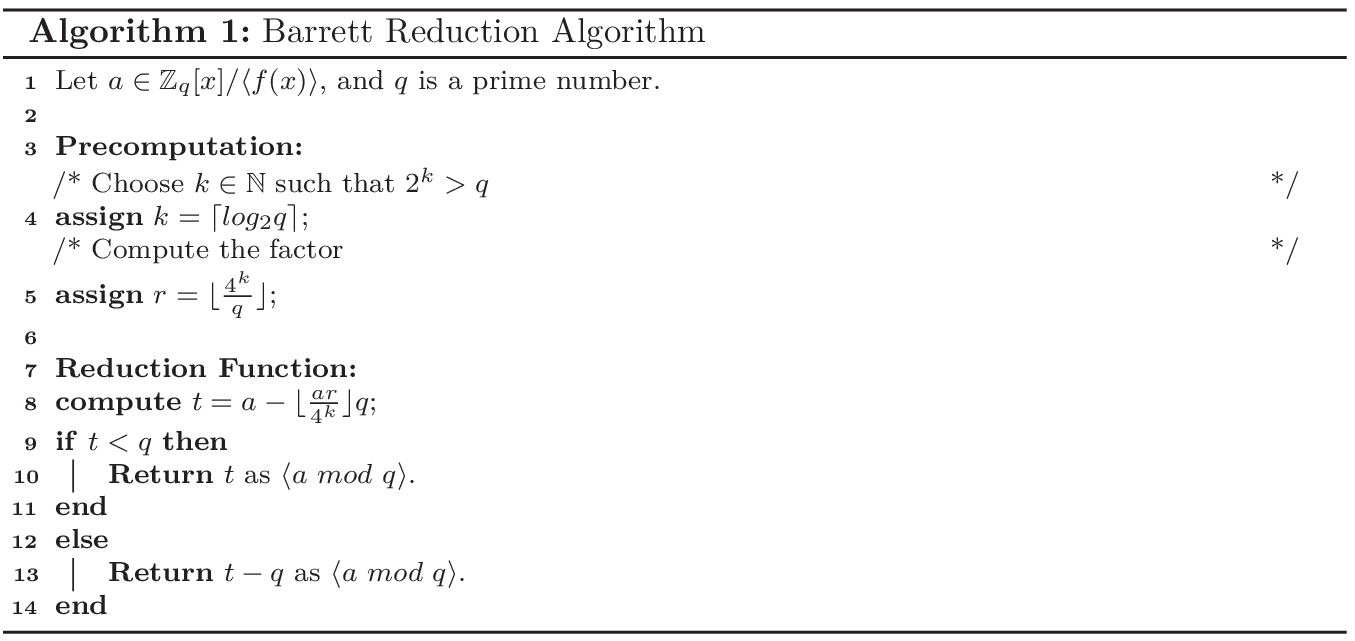} 
\end{center}
	\label{alg:BR}
\end{figure}

\begin{figure}[ht]
\begin{center}
	\includegraphics[width=0.95\columnwidth]{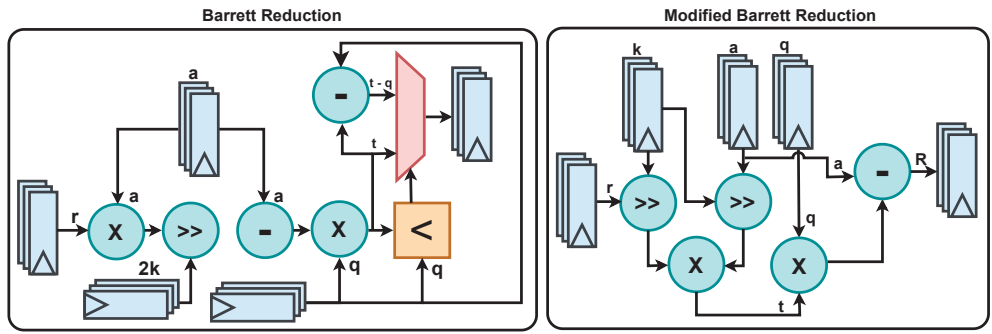} 
\end{center}
	\caption{Hardware implementation of Barrett reduction and Modified Barrett reduction.} 
	\label{fig:MR}
\end{figure}

Since the modulus $q$ is known in advance and the factor $r$ depends only on this modulus, it can be precomputed and stored. Then, the reduction function requires only computing the remainder value, $t$. While computing the $t$ value, a division operation by $4^k$, being power-of-2, can be performed using the right-shift operation. Hence, the entire computation reduces to just two multiplications, one right-shift, and one subtraction operation. Furthermore, the computation is performed in one step and, thus, is performed in constant time. The hardware implementation circuit is as shown in Figure \ref{fig:MR}.

For our hardware-based Barrett reduction implementation, we specifically included some additional optimizations. One such optimization is a careful bit width analysis. Say that the modulus requires exactly $k$ bits, then the product $\lfloor \frac{ar}{4^k} \rfloor q$ fits in $2k$ bits. We also observed that the computed values $t$ do not need more than $m + 1$ bits. The advantage of this observation is that we can safely ignore the upper $m - 1$ bits of the product. This in turn reduces the size of the registers to $m + 1$ bits while performing computations. 

\subsubsection{Modified Barrett reduction}
Hasenplaugh et al. \cite{HG2007} introduced an iterative folding method as a modification to the Barrett reduction method. This method not only reduces the number of required multiplications via an increased number of precomputations, but also reduces the bit width of the operations performed. We modify their proposed approach and propose Algorithm \textbf{2}, which computes modulo reduction in a single fold.

\begin{figure}[ht]
\begin{center}
	\includegraphics[width=0.99\columnwidth]{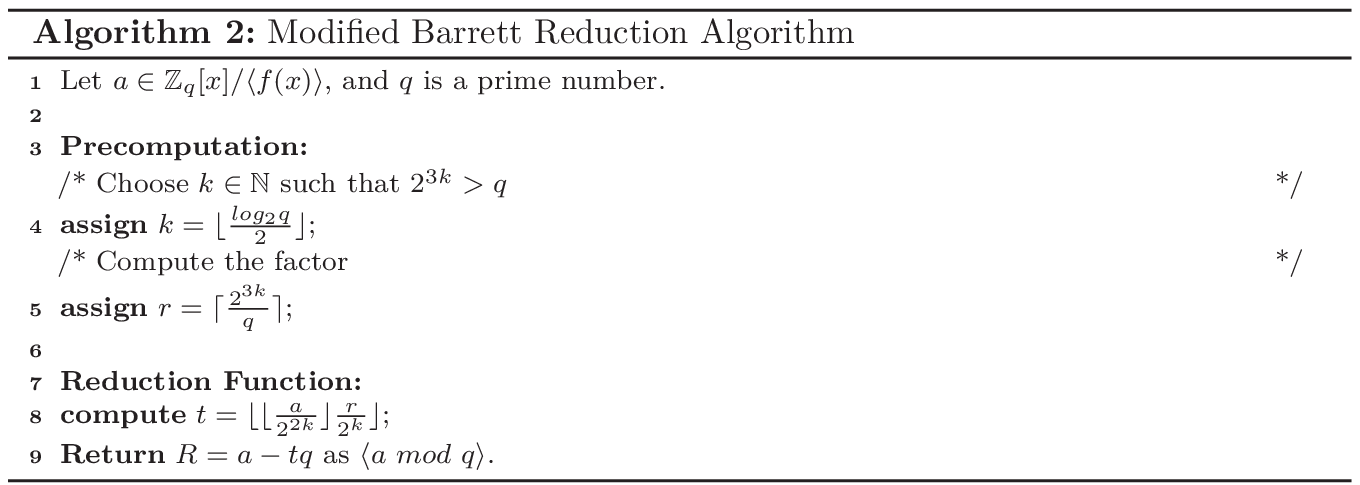} 
\end{center}
	\label{alg:MBR}
\end{figure}

When compared to Barrett reduction, the proposed algorithm precomputes $k$ with half the bit width and $r$ with one third the bit width. This significantly reduces the multiplication bit-width while performing actual computations after the coefficient integers are known. Moreover, we were able to get rid of an additional check, $t < q$, which is required in case of Barrett reduction. Hence, the result for modulo reduction is available with minimal operations using minimal bit width. The modified Barrett reduction's implementation is shown in Figure \ref{fig:MR}.

\subsection{Polynomial Multiplication using NTT} 
\label{PolyMul}
Polynomial multiplication is the most performed operation in homomorphic encryption and has the highest implementation complexity. Therefore, the latency of the polynomial multiplication module will govern the efficiency of the entire implementation. Hence, it is critical to design an efficient polynomial multiplication module. A conventional approach to implement a polynomial multiplier is to use convolution method. However, this approach is expensive to implement in hardware, as it requires performing $O(n^2)$ multiplications for a degree $n$ polynomial. This complexity can be reduced to $O(n \log_2 n)$ multiplications instead by using NTT combined with negative wrapped convolution to perform polynomial multiplication. We leverage the NTT-based multiplication algorithm proposed by Chen et al. \cite{CH2015} in our implementation. The steps involved in this algorithm are described in Algorithm \textbf{3}.

\begin{figure}[ht]
\begin{center}
	\includegraphics[width=0.99\columnwidth]{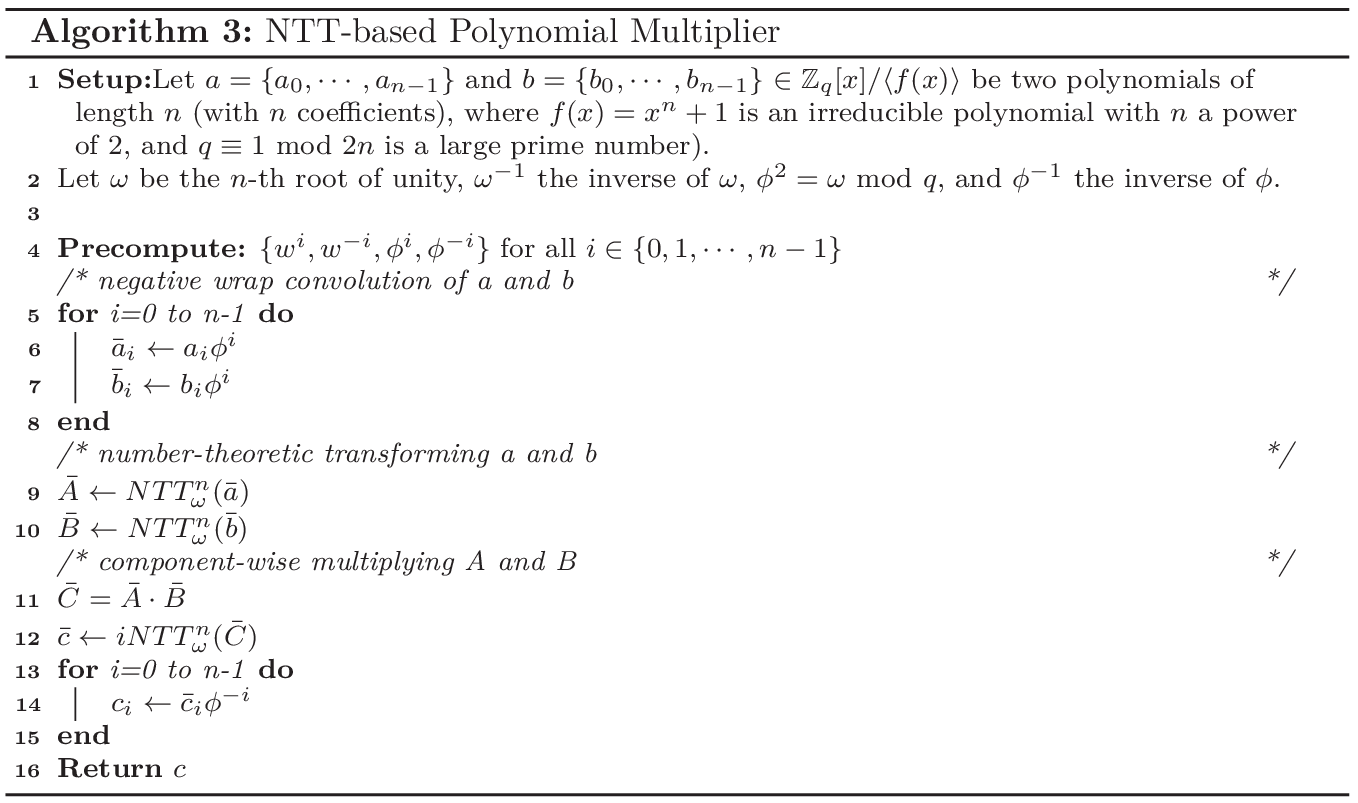} 
\end{center}
	\label{alg:MUL}
\end{figure}

\subsubsection{Number Theoretic Transform}
\label{NTT}
A generalization of the Fast Fourier Transform (FFT) over a finite ring $R_q = R/\langle q \rangle = \mathbb{Z}_q[x]/\langle f(x) \rangle$ is represented by Number Theoretic Transform (NTT). The equation of the NTT is as follows:
    \begin{equation} \label{eq: NTT}
		X_i = \sum_{k = 0}^{n-1} x_k \cdot \omega^{ik} 
	\end{equation}

where $\omega$ is the $n^{th}$ root of unity in the corresponding polynomial field and for a ring ${R_q}$, where $q$ is a prime number, the $n^{th}$ root of unity $\omega$ must satisfy two conditions: 
\begin{enumerate}
    \item $\omega^{n}$ = 1 mod $q$, \vspace{-0.1in}
    \item The period of $\omega^{i}$ for $i \in \{0, 1, 2, \cdots n-1\} $ is exactly $n$. 
\end{enumerate}

One of the efficient ways to compute $\omega$ is by using the following approach: 
\begin{enumerate}
    \item First compute the primitive root of $q$, which must satisfy: \vspace{-0.1in}
        \begin{itemize}
            \item $\alpha^{q-1}$ = 1 mod $q$
            \item The period of $\alpha^{i}$ for $i \in \{0, 1, 2, \cdots q-1\} $ is exactly $q-1$.
        \end{itemize}
    \item And since $\omega^{n} \equiv \alpha^{q-1}$ mod $q$, we can compute: 
    	\begin{equation*}
	    	\omega = \alpha^{(q-1)/n} \text{ mod } q
    	\end{equation*}
    \item As a final step, verify that this $\omega$ meets both the conditions mentioned above.
\end{enumerate}

\begin{figure}[ht]
\begin{center}
	\includegraphics[width=0.99\columnwidth]{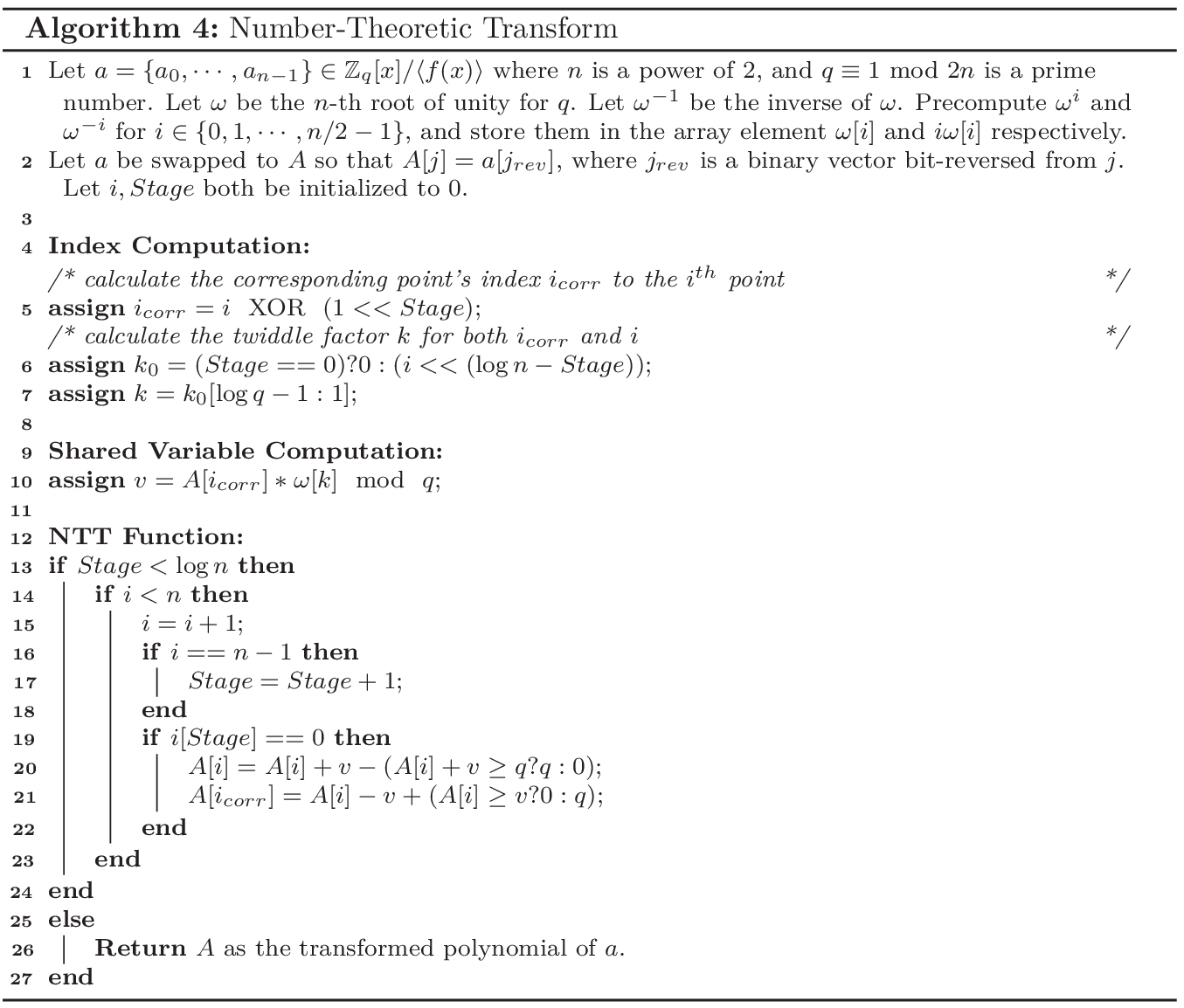} 
\end{center}
	\label{alg:NTT}
\end{figure}

Applying inverse NTT (iNTT) is straight forward and can be performed using the existing NTT module by replacing $\omega$ with $\omega^{-1}$, where $\omega^{-1}$ = $\omega^{n-1}$ mod $q$. iNTT computation also requires computing the inverse of $n$, which can be computed as $n^{-1} \cdot n$ = 1 mod $q$. 

\begin{figure}[ht]
	\begin{center}
		\includegraphics[width=5.3in]{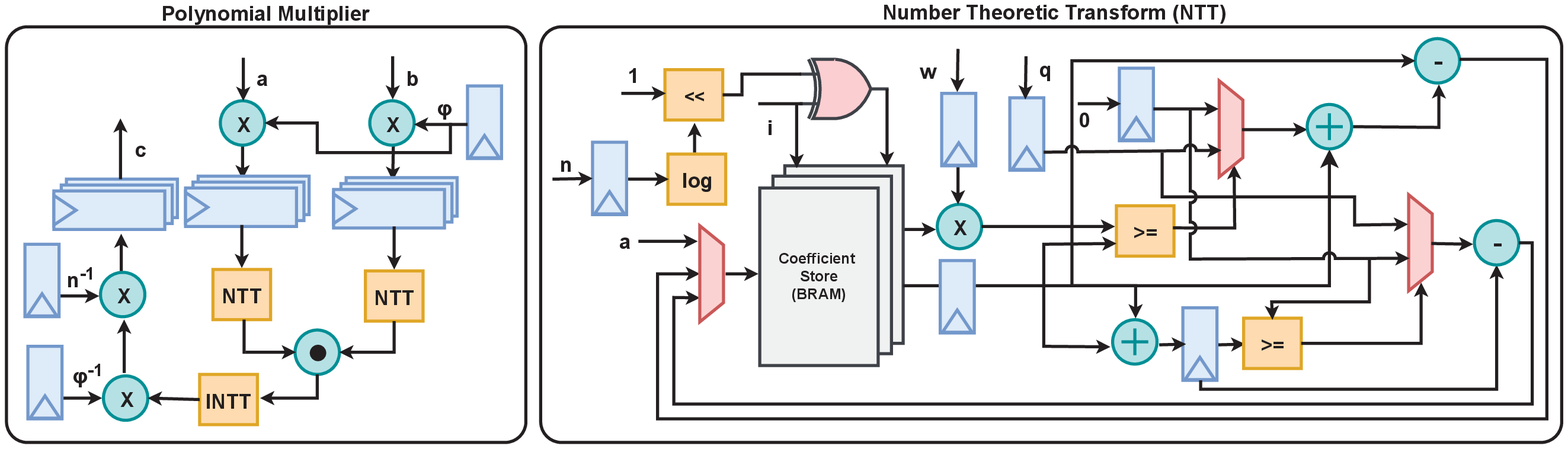}
	\end{center}
	\caption{Polynomial Multiplication and Operation within NTT}
	\label{fig:PolyMulNTT}
\end{figure}

Although there exist hardware implementations of NTT, they are quite expensive because of the way they compute the indices of the points and the corresponding $w^i$. Investing a large number of multiplications and divisions for these computations may not be an issue with software implementation, however they lead to a higher resource consumption in the hardware counterpart. Therefore, in our implementation of the NTT algorithm, Algorithm \ref{alg:NTT}, we perform the indices computation using only shift and xor operations. The benefit of doing so is that the shift and xor are not only inexpensive to implement but they conveniently replace the large multiplication and division circuits. By leveraging this highly optimized NTT implementation, we implement the fast polynomial multiplication algorithm (Algorithm \textbf{4}) very efficiently. Figure \ref{fig:PolyMulNTT} shows a high level circuit for polynomial multiplication and the operations within the NTT block.           

\subsection{Polynomial Addition} 
\label{PolyAdd}
Polynomial addition is the second most frequently used operation after polynomial multiplication. The schematic of the hardware implementation for polynomial addition is shown in Figure \ref{fig:submodules}. The implementation performs a component-wise addition operation on the coefficients of the polynomial. Note that the results are wrapped either within small modulus $q$ or large modulus $Q$ depending on which main module is utilizing this submodule to perform polynomial addition.  

\subsection{Scalar Multiplication} 
\label{ScMul}
\begin{figure}[ht]
    \centering
	\includegraphics[width=1\columnwidth]{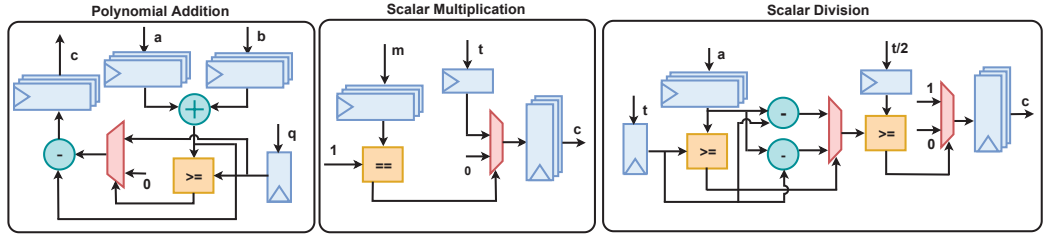} 
	\caption{Polynomial Addition, Scalar Multiplication, and Scalar Division submodules.} 
	\label{fig:submodules}
\end{figure}

Since the message space is binary, a conditional assignment operator can be used to implement the scalar multiplication operation. As shown in Figure \ref{fig:submodules}, $m$, the plaintext message, is an $n$-bit vector. Thus, computing $tm$ essentially requires choosing $t$ or $0$ according to each bit of $m$. Since we avoid performing actual multiplication operations, hardware cost is greatly reduced.

\subsection{Scalar Division to the Nearest Binary Integer} 
\label{ScDiv}
\vspace{-0.05in}
The scheme parameter $t = \lfloor \frac{Q}{2} \rfloor$ value is already published, and hence, it is known. From the equations \ref{equ:AddDec} and \ref{equ:MulDec} in the scheme, if we denote $u = (c_0 + s \cdot c_1)$, then to decrypt the message $m$ correctly, all we need to do is to compute $m = \lceil \frac{u}{t} \rfloor$. The nearest binary integer equivalent of $\lceil \frac{u}{t} \rfloor$ can be computed by measuring the distance between $u$ and $t$ as $(\text{Absolute}(u - t) < \frac{t}{2})\ ?\ 1 : 0$. If this distance is larger than half of $t$, it indicates that $u$ and $t$ are far from each other, and thus, the nearest integer of the quotient $\frac{u}{t}$ must be $0$. But if this distance is less than half of $t$, then the nearest integer of the quotient $\frac{u}{t}$ must be $1$. Thus, we implement the scalar division hardware circuit as shown in Figure \ref{fig:submodules}, without using any hardware division circuit.

\subsection{Chinese Remainder Theorem}
The Chinese remainder theorem, CRT \cite{KM2007}, states that if we know the residue of an integer, $a$ modulo two primes $q_1$ and $q_2$, it is possible to reconstruct $<a>_{q_1q_2}$ as follows. Let $<a>_{q_1}$ = $a_1$ and $<a>_{q_2}$ = $a_2$, then the value of $a \pmod Q$, where $Q=q_1\cdot q_2$, can be found by 
\begin{equation}
    a = <q_1t_1a_2 + q_2t_2a_1>Q
    \label{eq:CRT}
\end{equation}

where $t_1$ is the multiplicative inverse of $q_1 \pmod q_2$ and $t_2$ is the multiplicative inverse of $q_2 \pmod q_1$. This is feasible as the inverses $t_1$ and $t_2$ always exist, since $q_1$ and $q_2$ are coprime. Mathematically, $<a>_{q_1q_2}$ can also be represented by a set of congruent equations as follows:
\begin{align}
    \begin{split}
        a \equiv a_1 \pmod {q_1} \\
        a \equiv a_2 \pmod {q_2}
    \end{split}
\end{align}

Using the CRT, we combine all the small integers back into one large integer, so as to generate the required final result. A naive approach to implement CRT would compute pairwise multiplicative inverse or modulo inverse for two given moduli and then use equation \ref{eq:CRT} to merge values of $a_1$ and $a_2$ to get $a$. This process can be carried out recursively until the final coefficient value is obtained.

\begin{figure}[ht]
\begin{center}
	\includegraphics[width=0.99\columnwidth]{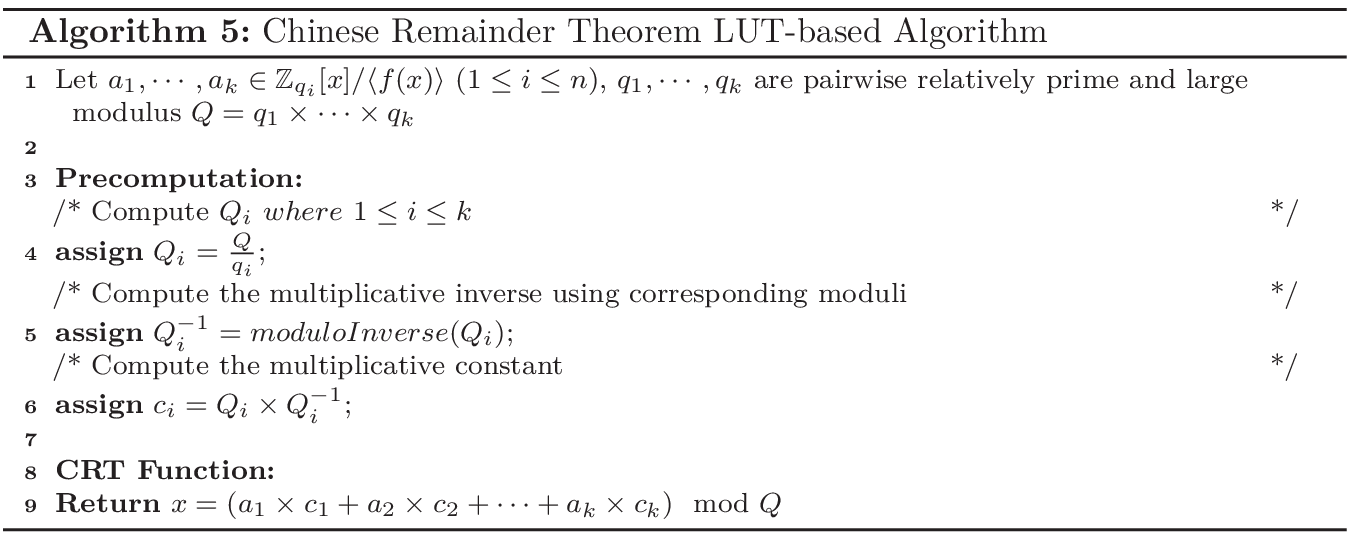} 
\end{center}
	\label{alg:CRTLuT}
\end{figure}

\begin{figure}[ht]
\begin{center}
	\includegraphics[width=0.45\columnwidth]{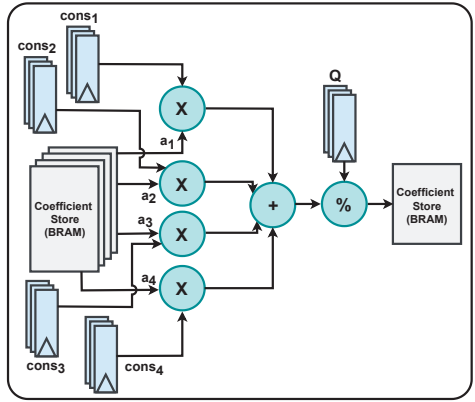} 
\end{center}
	\caption{Hardware implementation of CRT.} 
	\label{fig:CRT}
\end{figure}

The problem with this approach is that computing the multiplicative inverse at runtime increases latency. Therefore, a better approach is to precompute the modulo inverse values since all the moduli are known in advance. And then the actual computation reduces to just a single step, which can be performed in one clock cycle. The precomputation and computation steps involved are shown in Algorithm \textbf{5}. We call this approach LUT-based, since the precomputated values are stored in LUTs; its hardware implementation circuit is shown in Figure \ref{fig:CRT}. The hardware cost for LUT-based CRT can be further optimized by breaking down the single step multiplication and addition operation into various steps. This will enable the reuse of multipliers and adders during these steps. We leave this as future work for now.

\subsection{Modulo Inverse}
A modulo inverse or multiplicative inverse is the main computation involved in CRT. Moreover, while working with homomorphic encryption, due to large parameters, the amount of storage available can be a concern. Thus, instead of using LUT-based CRT, we may need to use a regular CRT implementation with the modulo inverse computed on the fly.   

The multiplicative inverse of $a \pmod q$ exists if and only if $a$ and $q$ are relatively prime (i.e., if $gcd(a, q) = 1$). Given two integers $a$ and $q$, the modulo inverse is defined by an integer $p$ such that
\begin{equation}
    a \cdot p \equiv 1 \pmod q
    \label{eq:MI}
\end{equation}

Here, the value of $p$ should be in ${0, 1, 2, \ldots q-1}$, i.e., in the range of integer modulo $q$. In our case, we need to compute the multiplicative inverse between pairwise moduli, $q_i$. There are two primary algorithms used in computing a modulo inverse. We discuss both algorithms in detail next.

\subsubsection{Fermat’s Little Theorem}
Fermat's little theorem \cite{VE2016} is typically used to simplify the process of modular exponentiation. But since we know $q$ is prime, we can also use Fermats’s little theorem to find the modulo inverse. According to this theorem, we can rewrite equation \ref{eq:MI} as follows:
\begin{equation}
    a^{q-1} \equiv 1 \pmod q
    \label{eq:FL1}
\end{equation}

If we multiply both sides of this equation with $a^{-1}$, we get
\begin{equation}
    a^{-1} \equiv a^{q-2} \pmod q
    \label{eq:FL2}
\end{equation}

Equation \ref{eq:FL2} is the only computation carried out by this theorem to get the value of $a^{-1}$ and this is what is shown in Algorithm \textbf{6}. The algorithm has a time complexity of $O(log^2 q)$. For our hardware-based implementation, we precompute the power factors from $1$ to $q-2$ to save the computation cost. This not only speeds up computation but also significantly reduces the hardware cost. The hardware implementation is shown in Figure \ref{fig:MI}.       

\begin{figure}[ht]
\begin{center}
	\includegraphics[width=0.99\columnwidth]{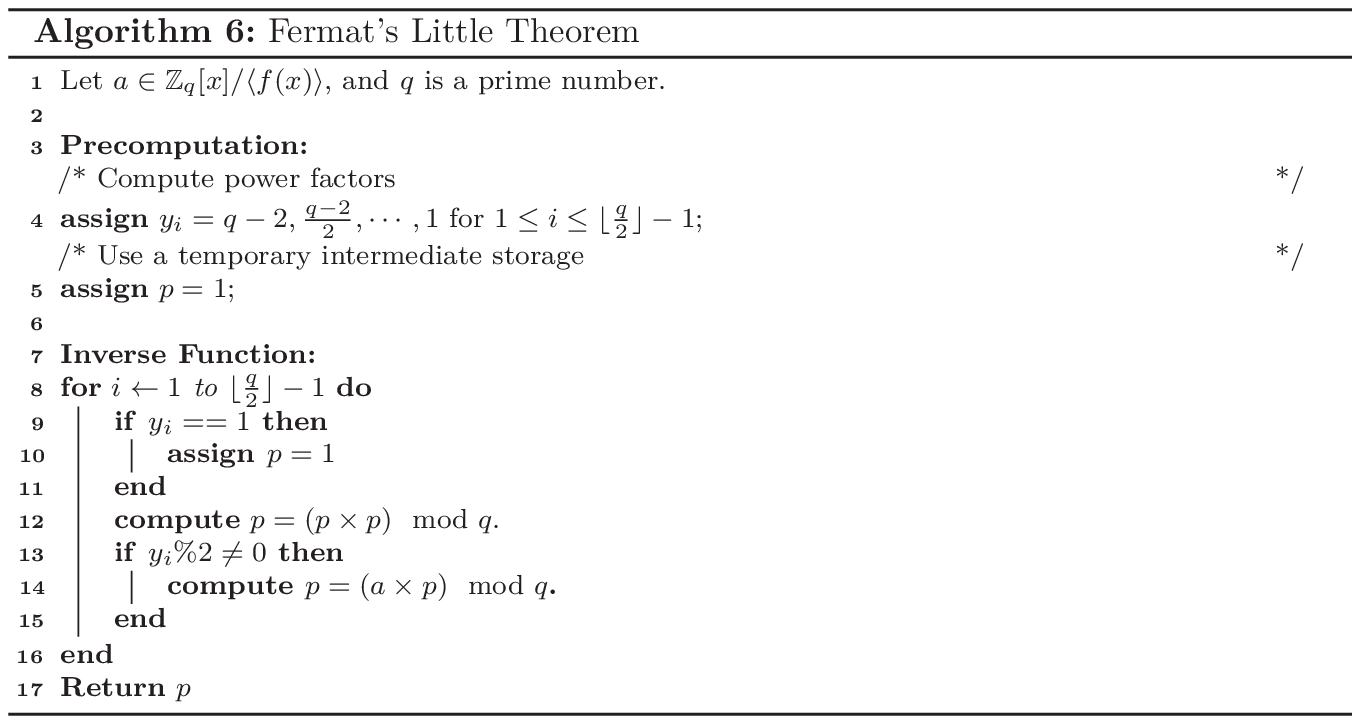} 
\end{center}
	\label{alg:FLT}
\end{figure}

\subsubsection{Extended Euclidean Algorithm}
The extended Euclidean algorithm \cite{VE2016} is an extension to the classic Euclidean algorithm that is used for finding the greatest common divisor (GCD). According to this algorithm, if $a$ and $q$ are relatively prime, there exist integers $x$ and $y$ such that $ax + qy = 1$, and such integers may be found using the Euclidean algorithm. Considering this equation modulo $q$, it follows that $ax = 1$; i.e., $x=a^{-1} \pmod q$. The algorithm used for our implementation is as shown in Algorithm \textbf{7}. 

The algorithm works as follows. Given two integers $0 < a < q$, using the classic
Euclidean algorithm equations, one can compute $gcd(a,q) = r_j$, where $r_j$ is the remainder. In the classic
Euclidean algorithm, we start by dividing $q$ by $a$ (integer division with remainder), then repeatedly divide the previous divisor by the previous remainder until there is no remainder. The last remainder we divided by is the greatest common divisor. To avoid division operations, the classic Euclidean algorithm equations can be rewritten as follows:
\vspace{-0.1in}
\begin{align*}
    r_1 &= a - q \cdot x_1, \\
    r_2 &= a - r_1 \cdot x_2, \\
    r_3 &= r_1 - r_2 \cdot x_3, \\
    &\vdots\\
    r_j &= r_{j-2} - r_{j-1} \cdot x_j 
\end{align*}

Then, in the last of these equations, $r_j = r_{j-2} - r_{j-1} \cdot x_j$ , replace $r_{j-1}$ with its expression in terms of $r_{j-3}$ and $r_{j-2}$ from the equation immediately above it. Continue this process
successively, replacing $r_{j-2},r_{j-3},\ldots,$ until we obtain the final equation $r_j = ax + qy$,
with $x$ and $y$ integers. In our special case that $gcd(a,b) = 1$, the integer equation reads as $1 = ax + qy$ and therefore we deduce $1 \equiv ax \pmod q$ so that the residue of $x$ is the multiplicative inverse of $a \pmod q$. The time complexity of this algorithm is $O(log(min(a,q)))$.

\begin{figure}[ht]
\begin{center}
	\includegraphics[width=0.99\columnwidth]{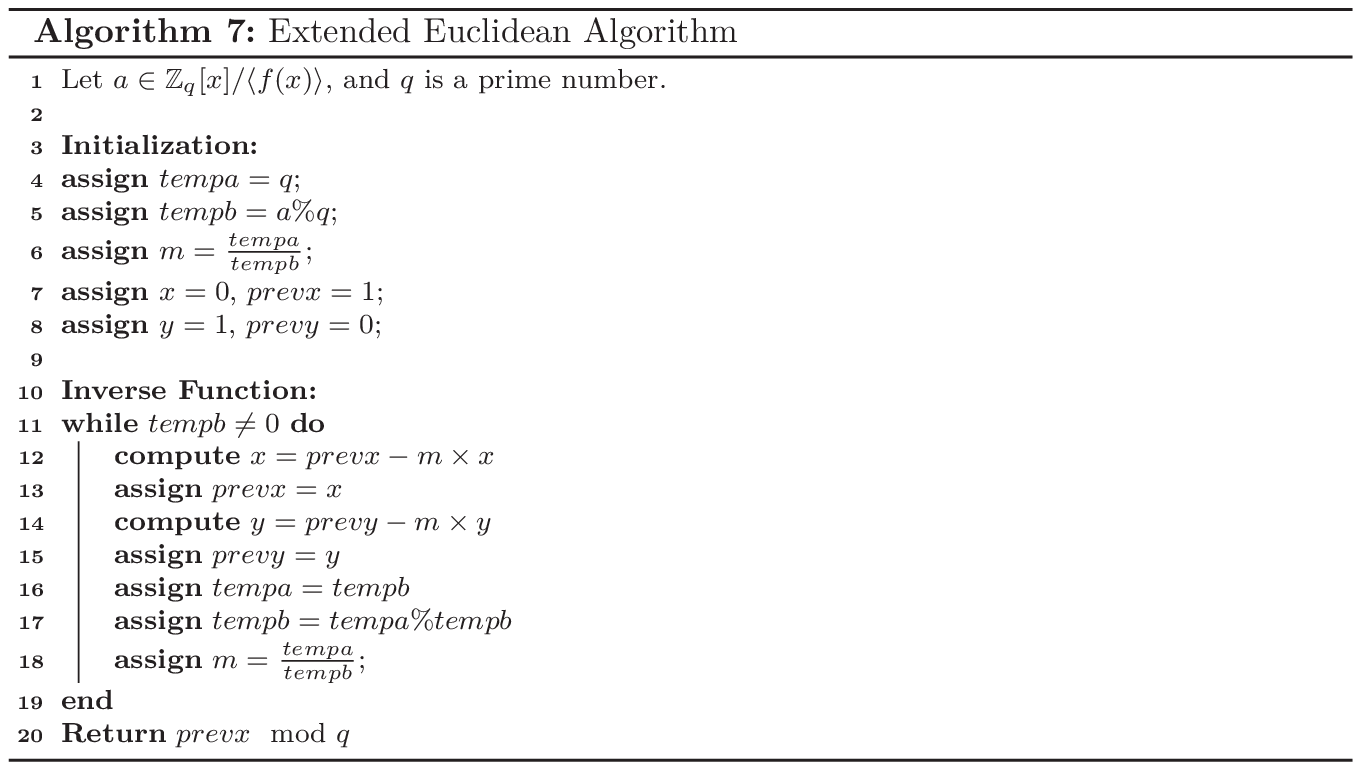} 
\end{center}
	\label{alg:EEA}
\end{figure}

\begin{figure}[ht]
\begin{center}
	\includegraphics[width=0.95\columnwidth]{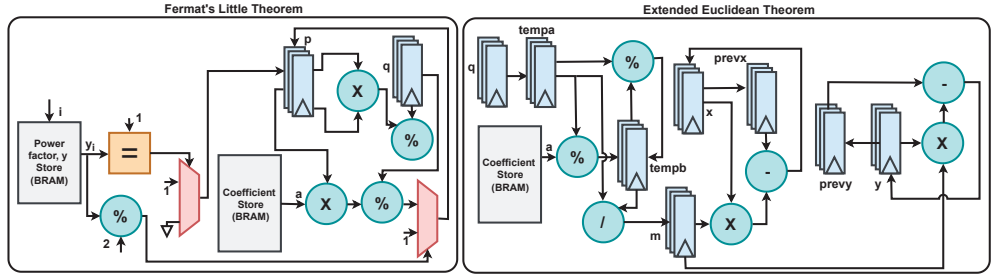} 
\end{center}
	\caption{Hardware implementation of Fermat's little and extended Euclidean theorem.} 
	\label{fig:MI}
\end{figure}

The hardware implementation is shown in Figure \ref{fig:MI}. The algorithm is simplified by removing unnecessary variables and computations to make it more suitable for hardware implementation. The implementation is done in an iterative fashion so that the input parameters gradually decrease while keeping the GCD of the parameters
unchanged. 
\subsection{Gaussian Noise Sampler}
\label{GNS}
The security of the RLWE-based encryption scheme is governed by small error samples generated from a Gaussian distribution. Hence, a Gaussian noise sampler lies at the core of maintaining the required security level. However, it is critical to select a sampling algorithm with a high sampling efficiency and throughput so that the key generation and the encryption operations, at the client side, still remain efficient. We leverage the implementation of a Ziggurat-based Gaussian noise sampler done by the authors in \cite{AP2020}. Due to space constraints we do not present the implementation details here but the interested readers can refer to the actual paper.      
\subsection{Relinearisation}
\label{relin}
\begin{figure}[ht]
\begin{center}
	\includegraphics[width=1\columnwidth]{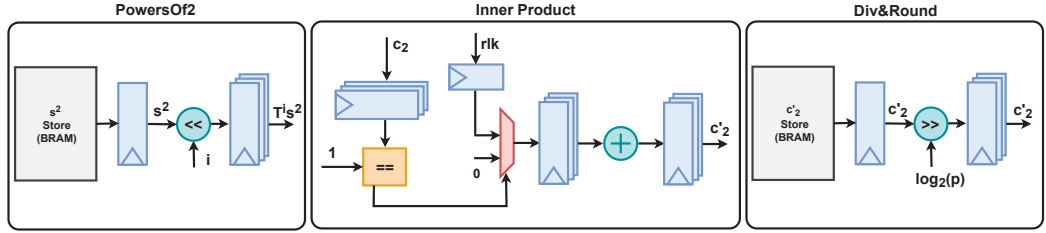} 
\end{center}
	\caption{PowersOf2, Inner Product, and Div\&Round submodules.} 
	\label{fig:Relin}
\end{figure}

We discuss relinearisation version $1$ implementation details first. In version $1$, the key generation will reuse most of the existing submodules except for the powers-of-2 computation. This operation is indicated by the PowersOf2 submodule in Figure \ref{fig:Relin}. The values of $T^i$ are not precomputed and stored to reduce the memory overhead. Instead, we take the vector $s^2$ and perform a left shift operation on all of the elements of this vector. The first set of left shift operations should be by $0$ bits to indicate $2^0$ multiplication, then by $1$ bit for $2^1$ multiplication, and so on until $2^\ell$ multiplications are performed. 

Note that we generate the relinearisation keys in $R_{q_1}, \ldots, R_{q_k}$ rather than $R_Q$. This facilitates the routing of the relinearisation keys correctly to the corresponding $q_i$ operation pipeline without the need to perform modular reductions. Although we perform $k$ times more operations, these operations are significantly faster. Additionally, the output of this submodule is arranged in such a fashion that the elements having the same index, from key $rlk[0]$, are treated as a single output. A similar output format holds true for $rlk[1]$ as well. This helps in faster indexing of the relinearisation keys while computing the inner product with $c_2$. The schematic of PowersOf2 submodule implementation is as shown in Figure \ref{fig:Relin}.

In the relinearisation version $1$ module in Figure \ref{fig:Relin}, the decomposition submodule's task represents the ciphertext $c_2$ at bit level, i.e., converting the coefficients from $R_q$ to $R_T$ and here $T = 2$. We know that in hardware bit-level operations can be performed readily, and hence, this operation becomes trivial. Therefore, we do not specifically provide an implementation of this submodule, and it is shown for completeness in Figure \ref{fig:SHE}. Next, we describe the implementation of the inner product submodule. To avoid performing actual multiplication operations, we leverage our scalar multiplication module (Section \ref{ScMul}) within this submodule, since $c_2$ is binary. Hence, a conditional operator does the work of multiplying elements of $c_2$ and relinearisation keys. We just need to use adders to compute the summation to finish the inner product computations. Implementation of this submodule is shown in Figure \ref{fig:Relin}.        

We will explore the relinearisation version $2$ submodules now. For key generation, we pick the largest $q_i$ from the moduli set, compute $q_{i}^3$ and then the immediate next power of two is set as the value of $p$. This $p$ is the scaling factor. Since we choose a power of $2$ as $p$, we can simply perform shift left operations to emulate the multiplication of $p$ with $s^2$ while generating the relinearisation keys. Note that $rlk[1]$ or $a$ is sampled from $R_{p \cdot q_i}$. Additionally, to maintain the required security, the error samples need to be generated from a different noise sampler. Hence, a second instance of the noise sampler is used here with the required parameter settings. The rest of the submodules are as previously discussed. 

While performing the relinearisation operation in version $2$, the ciphertext needs to be scaled down. This task is accomplished by using the Div\&Round submodule shown in Figure \ref{fig:Relin}. As the scale factor $p$ is a power of $2$, division operations can be avoided, and shift right operations can be performed instead. Most other existing implementations (both software and hardware), precompute $\frac{1}{p}$, round it down, and perform multiplication operations instead of division. There are two disadvantages to this approach. First rounding leads to loss of precision, generating approximate results and magnifying the errors in decryption as the levels of operations increase. Second, even though the expensive division operations are avoided, multiplications are still costly, requiring large multipliers which are not only expensive but also lead to a lower operating frequency. Figure \ref{fig:Relin} shows the Div\&Round submodule circuit.

\section{Performance Evaluation}
\label{Perf}
We evaluate the performance of all the design implementations through synthesis on a Xilinx Zynq-7000 family xc7z020clg400-1 FPGA. The tool used for synthesis is ISE design suite 14.7, with all designs implemented in Verilog 2001. To generate synthesis results, the input coefficient bit width is considered $1200$ bits and there are $40$ coprime moduli, $q$, having $30$ bits each. The degree of the polynomial is taken as $1024$.

\subsection{Hardware Cost and Latency}
We start by discussing the hardware cost and latency of the individual operations listed in Table \ref{table:HC} and \ref{table:LF}. The hardware cost depends on the size of each coefficient (either $1200$ bits or $30$ bits) and the number of portions into which a $1200$-bit coefficient is divided. The latency computation factors in the number of portions, $k$, or the polynomial degree, $n$ as required by the implementation of a module. That is why in Table \ref{table:LF}, the latencies are represented as a factor of $k$ or $n$.

\begin{table}[h!]
\caption{Hardware cost of the individual operations.}
\begin{center}
  \begin{tabular}{ | l | c | c | c | c |} 
    \hline
    Operation & LUT Slices & Registers & DSP & BRAM \\    
    \hline
    \hline
    Mod Operator (\%) & 798 & 0 & 0 & 0 \\
    \hline
    Barrett Reduction & 71 & 0 & 0 & 0 \\
    \hline
    Modified Barrett Reduction & 23 & 0 & 3 & 0 \\
    \hline
    \hline
    RNS (serial) & 7592 & 90 & 0 & 1.5 \\
    \hline
    RNS (serial modified) & 145 & 56 & 3 & 1.5 \\
    \hline
    RNS (parallel) & 88353 & 1242 & 0 & 2 \\
    \hline
    RNS (parallel modified) & 133 & 86 & 3 & 2 \\
    \hline
    \hline
    CRT & 3883 & 2408 & 20 & 6 \\
    \hline
    CRT (LUT-based) & 1274 & 301 & 4 & 6 \\
    \hline
    \hline
    Modulo Inverse (Fermat's little) & 1889 & 120 & 14 & 1 \\
    \hline
    Modulo Inverse (Extended Euclidean) & 3993 & 154 & 3 & 1 \\
    \hline
    \hline
    NTT & 6188 & 1291 & 0 & 3 \\
    \hline
    NTT-based Polynomial Multiplication & 8261 & 162 & 30 & 6 \\
    \hline
    Polynomial Addition & 1185 & 56 & 0 & 1 \\
    \hline
    Scalar Multiplication & 118 & 10 & 0 & 3 \\
    \hline
    Scalar Division to nearest integer & 672 & 14 & 0 & 3 \\
    \hline
    \hline
    PowersOf2 & 113 & 20 & 0 & 1 \\
    \hline
    Inner Product & 8961 & 796 & 0 & 3 \\
    \hline
    Div\&Round & 0 & 30 & 0 & 1 \\
    \hline
\end{tabular}
\label{table:HC}
\end{center}
\end{table}

The built in Verilog $mod$ operator is very expensive and takes about $800$ LUTs to perform a $30$-bit modulo reduction. For the same bit width, the classic Barrett reduction reduces the hardware cost by almost $11$ times, and our proposed modified Barrett reduction reduces the cost by about $35$ times. Since modular reduction is performed very frequently and used by all the modules, using the modified Barrett reduction method substantially reduces the hardware resources required for implementing the entire homomorphic encryption scheme. Note that the latency of the modular reduction module is not shown, as the implementation comprises combinational logic only. We observe that the RNS parallel implementation utilizes about $12$ times more LUT slices, when compared to the serial implementation, while the latency of the serial implementation is about $40$ times higher than the parallel one. The RNS serial and parallel modified implementations listed in the table use the modified Barrett reduction to perform the modulo reduction operation, instead of the inbuilt Verilog $mod$ operator. The hardware resource utilization is dramatically reduced by this modification, however, the latency remains the same. 

\begin{table}[h!]
  \caption{Latency and frequency of the individual operations.}
  \begin{center}
  \begin{tabular}{ | l | c | c |} 
    \hline
    Operation & Latency (clock cycles) & Frequency (MHz) \\    
    \hline
    RNS (serial) & 120$n$ & 260.5 \\
    \hline
    RNS (serial modified) & 120$n$ & 265.3 \\
    \hline
    RNS (parallel) & 3$n$ & 314.1 \\
    \hline
    RNS (parallel modified) & 3$n$ & 316.4 \\
    \hline
    \hline
    CRT & 1404$n$ & 132.4 \\
    \hline
    CRT (LUT-based) & 156$n$ & 134.5 \\
    \hline
    \hline
    Modulo Inverse (Fermat's little) & 3240$n$ & 113.7 \\
    \hline
    Modulo Inverse (Extended Euclidean) & 360$n$ & 117.4 \\
    \hline
    \hline
    NTT & 10240$k$ & 218.1 \\
    \hline
    NTT-based Polynomial Multiplication & 20480$k$ & 121.9 \\
    \hline
    Polynomial Addition & 3072$k$ & 144.4 \\
    \hline
    Scalar Multiplication & 2048$k$ & 243.8 \\
    \hline
    Scalar Division to nearest integer & 2048$k$ & 129.5 \\
    \hline
    \hline
    PowersOf2 & 163840 & 349.1 \\
    \hline
    Inner Product & 153600 & 124.3 \\
    \hline
    Div\&Round & $k$ & 204.3 \\
    \hline
  \end{tabular}
  \label{table:LF}
  \end{center}
\end{table}

The regular CRT implementation is about $3$ times more expensive than a LUT-based CRT. This difference is because regular CRT spends a lot of hardware resources for computing the multiplicative inverse, while the LUT-based CRT, with all its precomputations, not only requires fewer hardware resources but also performs computations in $9$ times fewer clock cycles. While computing multiplicative inverses, Fermat's little theorem facilitates a low hardware cost implementation, with about half the hardware cost as compared to the widely used extended Euclidean method. However, the extended Euclidean method performs computations about $9$ times faster. An NTT-based polynomial multiplication cuts down the latency from $n^2$ to $n\log n$. The polynomial addition, scalar multiplication, and scalar division submodules avoid the usage of modular reduction, multiplication and division operations respectively. Hence, these submodules are implemented using minimal hardware resources and have a low latency. For the rest of the other submodules involved in relinearisation, because of all the optimizations in implementation, we observe a low hardware cost and latency.   

\subsection{Hardware library vs Software library Speedup}
Table \ref{table:HET} provides the time, in clock cycles, for computing various homomorphic encryption operations. We represent time in clock cycles due to frequency difference between FPGA and general-purpose CPU. 

\begin{table}[h!]
  \caption{Time required for homomorphic encryption operations.}
  \begin{center}
  \begin{tabular}{ | l | c | } 
    \hline
    Operation & Time (in clock cycles) \\    
    \hline
    \hline
    Homomorphic addition & $3072$ \\
    \hline
    Homomorphic multiplication &  $71338$ \\
    \hline
    \hline
    Relinearisation KeyGen (version 1) & $86698$ \\
    \hline
    Relinearisation (version 1) & $18432$ \\
    \hline
    \hline
    Relinearisation KeyGen (version 2) & $72362$ \\
    \hline
    Relinearisation (version 2) & $112298$ \\
    \hline
    \hline
    RNS + CRT &  $23259$ \\
    \hline
    Encryption &  $75434$ \\
    \hline
    Decryption (Degree-1) &  $73386$ \\
    \hline
    Decryption (Degree-2) &  $141653$ \\
    \hline
  \end{tabular}
  \label{table:HET}
  \end{center}
\end{table}

As seen in the table, a single HE addition is about $23\times$ faster than a HE multiplication. Moreover, if one has to choose between relinearisation version $1$ and $2$, then version $1$ would be the unanimous choice, as it is about $6 \times$ faster than version $2$, even when key generation takes almost the same time. It is worth noting that, although the table lists the relinearisation key generation time for both versions, these keys can be precomputed, and hence, the time required for key generation need not be included in the overall time required for HE operations. Based on the parameters that are used in the implementation, we can evaluate a circuit of depth $56$, with relinearisation performed after every multiplication operation. Therefore, using the data from Table \ref{table:HET}, we can compute the number of clock cycles required for this entire set of operations. We observe that when the circuit evaluation is done with relinearisation version $1$, the total cycles required are $5,031,984$, while evaluation done with relinearisation version $2$ takes $10,194,651$ cycles. 

\begin{table}[h!]
  \caption{Hardware speedup for homomorphic encryption operations.}
  \begin{center}
  \begin{tabular}{ | l| l | l| l| } 
    \hline
    Operation & \multicolumn{1}{|p{3.4cm}|}{\centering Palisade library \\(Time in clock cycles)} & \multicolumn{1}{|p{3.5cm}|}{\centering Our Hardware library \\(Time in clock cycles)} & Speedup \\    
    \hline
    \hline
    Encryption & 119700000 & 75434 & $1500\times$ \\
    \hline
    Homomorphic mult. & 299729520 & 71338 & $4200\times$ \\
    \hline
    Homomorphic add. & 9070884 & 3072 & $2950\times$ \\
    \hline
    Decryption & 22400640 & 73386 & $300\times$ \\
    \hline
  \end{tabular}
  \label{table:HWS}
  \end{center}
\end{table}

Next, we present the speed up obtained by the hardware accelerator designed using the modules in our hardware library in comparison to its software counterpart. For this comparison, we recorded the number of clock cycles required for encryption, HE multiplication, HE addition and decryption in the Palisade software library using same underlying scheme with RNS implementation using same parameters. Table \ref{table:HWS} lists the observed speedup. This evaluation assumes we utilize the maximum possible resources available on the FPGA that we used for our evaluation. There is a scope to further enhance the speedup using more hardware resources. 

\begin{table}[h!]
  \caption{Hardware speedup for Logistic Regression prediction.}
  \begin{center}
  \begin{tabular}{ | l | c | } 
    \hline
    Operation & Speedup (in clock cycles) \\    
    \hline
    Logistic Regression prediction & $2650\times$ \\
    \hline
  \end{tabular}
  \label{table:HWSLR}
  \end{center}
\end{table}

Finally, we evaluate the time required to make a prediction using logistic regression, a common tool used in machine learning for binary classification problems. Making predictions using logistic regression model with $x_0,x_1,...,x_n$ features require computing the logistic regression equation, $Y = {{e^{X}}/{1 + e^{X}}}$, where $X = \sum^{n}_{i=0}b_ix_i$. We first compute $X$ and then use the value in the Remez algorithm (an iterative minimax approximation algorithm \cite{FW1965}) equation, $Y(X) = -0.004X^3 + 0.197X + 0.5$  to compute the probability. The use of Remez algorithm equation helps avoid log computation which is required in the logistic regression equation. However, these equations require working over floating point numbers and the FV scheme does not support the encryption of floating point numbers. Hence, we scale them to integers and perform fixed point operations instead. We observe that our hardware accelerator requires $470,064$ clock cycles to perform one logistic regression prediction using homomorphic encryption, providing a speedup of around $2650\times$ over a software-based prediction as mentioned in Table \ref{table:HWSLR}.

\section{Conclusion}
\label{con}
We presented a fast hardware arithmetic hardware library with a focus on accelerating the key arithmetic operations involved in RLWE-based somewhat homomorphic encryption. For all of these operations, we include a hardware cost efficient serial implementation and a fast parallel implementation in the library. We also presented a modular and hierarchical implementation of a hardware accelerator using the modules of the proposed arithmetic library to demonstrate the speedup achievable in hardware. The parameterized design implementation approach of the modules and the hardware accelerator provides the flexibility to extend use of the modules for other schemes, such as BGV, and the accelerator for many applications, especially in the FPGA-centric cloud computing environment. Evaluation of the implementation shows that a speed up of about $4200\times$ and $2950\times$ for evaluating homomorphic multiplication and addition respectively is achievable in hardware when compared to software implementation.

As future work, we would like to optimize and implement the arithmetic operations involved in bootstrapping as well. The bootstrap operation is one of the key functions in achieving fully homomorphic encryption, but it remains very expensive to perform. Optimizing the bootstrap operation will render it more practical to use. We are also actively working on integrating other RLWE-based homomorphic encryption schemes, like BGV, into our library so as to leverage inherent advantages that these schemes offer. Once we have the required operations and schemes implemented, we will open-source the arithmetic library and FPGA design examples.

\bibliographystyle{alpha}
\bibliography{references}

\newcommand{\etalchar}[1]{$^{#1}$}
\begin{thebibliography}{AMBG{\etalchar{+}}16}

\bibitem[ABK]{AP2020}
Rashmi Agrawal, Lake Bu, and Michel~A Kinsy.
\newblock A post-quantum secure discrete gaussian noise sampler.
\newblock {\em 2020 IEEE International Symposium on Hardware Oriented Security
  and Trust (HOST)}.

\bibitem[AMBG{\etalchar{+}}16]{AN2016}
Carlos Aguilar-Melchor, Joris Barrier, Serge Guelton, Adrien Guinet,
  Marc-Olivier Killijian, and Tancrede Lepoint.
\newblock Nfllib: Ntt-based fast lattice library.
\newblock In {\em Cryptographers’ Track at the RSA Conference}, pages
  341--356. Springer, 2016.

\bibitem[Bar87]{BP1986}
Paul Barrett.
\newblock Implementing the rivest shamir and adleman public key encryption
  algorithm on a standard digital signal processor.
\newblock In Andrew~M. Odlyzko, editor, {\em Advances in Cryptology --- CRYPTO'
  86}, pages 311--323, Berlin, Heidelberg, 1987. Springer Berlin Heidelberg.

\bibitem[BFZY11]{BD2011}
Kang Bing, Liu Fu, Yun Zhuo, and Liang Yanlei.
\newblock Design of an internet of things-based smart home system.
\newblock In {\em 2011 2nd International Conference on Intelligent Control and
  Information Processing}, volume~2, pages 921--924. IEEE, 2011.

\bibitem[BGV14]{BL2014}
Zvika Brakerski, Craig Gentry, and Vinod Vaikuntanathan.
\newblock (leveled) fully homomorphic encryption without bootstrapping.
\newblock {\em ACM Transactions on Computation Theory (TOCT)}, 6(3):13, 2014.

\bibitem[CKKS17]{CH2017}
Jung~Hee Cheon, Andrey Kim, Miran Kim, and Yongsoo Song.
\newblock Homomorphic encryption for arithmetic of approximate numbers.
\newblock In {\em International Conference on the Theory and Application of
  Cryptology and Information Security}, pages 409--437. Springer, 2017.

\bibitem[CLP]{CE20}
Anamaria Costache, Kim Laine, and Rachel Player.
\newblock Evaluating the effectiveness of heuristic worst-case noise analysis
  in fhe.

\bibitem[CLP17]{CS2017}
Hao Chen, Kim Laine, and Rachel Player.
\newblock Simple encrypted arithmetic library-seal v2. 1.
\newblock In {\em International Conference on Financial Cryptography and Data
  Security}, pages 3--18. Springer, 2017.

\bibitem[CMV{\etalchar{+}}15]{CH2015}
Donald~Donglong Chen, Nele Mentens, Frederik Vercauteren, Sujoy~Sinha Roy,
  Ray~CC Cheung, Derek Pao, and Ingrid Verbauwhede.
\newblock High-speed polynomial multiplication architecture for ring-lwe and
  she cryptosystems.
\newblock {\em IEEE Transactions on Circuits and Systems I: Regular Papers},
  62(1):157--166, 2015.

\bibitem[DS15]{DC2015}
Wei Dai and Berk Sunar.
\newblock cuhe: A homomorphic encryption accelerator library.
\newblock In {\em International Conference on Cryptography and Information
  Security in the Balkans}, pages 169--186. Springer, 2015.

\bibitem[Fra65]{FW1965}
W.~Fraser.
\newblock A survey of methods of computing minimax and near-minimax polynomial
  approximations for functions of a single independent variable.
\newblock {\em J. ACM}, 12(3):295–314, July 1965.

\bibitem[FV12]{FS2012}
Junfeng Fan and Frederik Vercauteren.
\newblock Somewhat practical fully homomorphic encryption.
\newblock {\em IACR Cryptology ePrint Archive}, 2012:144, 2012.

\bibitem[G{\etalchar{+}}09]{GF2009}
Craig Gentry et~al.
\newblock Fully homomorphic encryption using ideal lattices.
\newblock In {\em Stoc}, volume~9, pages 169--178, 2009.

\bibitem[Gar59]{GR1959}
Harvey~L Garner.
\newblock The residue number system.
\newblock In {\em Papers presented at the the March 3-5, 1959, western joint
  computer conference}, pages 146--153. ACM, 1959.

\bibitem[GHS12]{GC2012}
Craig Gentry, Shai Halevi, and Nigel~P. Smart.
\newblock Homomorphic evaluation of the aes circuit.
\newblock In Reihaneh Safavi-Naini and Ran Canetti, editors, {\em Advances in
  Cryptology -- CRYPTO 2012}, pages 850--867, Berlin, Heidelberg, 2012.
  Springer Berlin Heidelberg.

\bibitem[Hay08]{HC2008}
Brian Hayes.
\newblock Cloud computing.
\newblock {\em Communications of the ACM}, 51(7):9--11, 2008.

\bibitem[HGG07]{HG2007}
William Hasenplaugh, Gunnar Gaubatz, and Vinodh Gopal.
\newblock Fast modular reduction.
\newblock In {\em Proceedings of the 18th IEEE Symposium on Computer
  Arithmetic}, ARITH '07, pages 225--229, Washington, DC, USA, 2007. IEEE
  Computer Society.

\bibitem[HS14]{HS2014}
Shai Halevi and Victor Shoup.
\newblock Algorithms in helib.
\newblock In Juan~A. Garay and Rosario Gennaro, editors, {\em Advances in
  Cryptology -- CRYPTO 2014}, pages 554--571, Berlin, Heidelberg, 2014.
  Springer Berlin Heidelberg.

\bibitem[KI07]{KM2007}
Victor~J Katz and Annette Imhausen.
\newblock {\em The Mathematics of Egypt, Mesopotamia, China, India, and Islam:
  A Sourcebook}.
\newblock Princeton University Press, 2007.

\bibitem[KIA11]{KC2011}
SO~Kuyoro, F~Ibikunle, and O~Awodele.
\newblock Cloud computing security issues and challenges.
\newblock {\em International Journal of Computer Networks (IJCN)},
  3(5):247--255, 2011.

\bibitem[Kim18]{KH2018}
Andrey Kim.
\newblock \uppercase{HEAAN}, 2018.

\bibitem[KJBB16]{KI2016}
Ravi~Kishore Kodali, Vishal Jain, Suvadeep Bose, and Lakshmi Boppana.
\newblock Iot based smart security and home automation system.
\newblock In {\em 2016 international conference on computing, communication and
  automation (ICCCA)}, pages 1286--1289. IEEE, 2016.

\bibitem[LT219]{LT2019}
Lattigo 1.3.0.
\newblock Online: \url{http://github.com/ldsec/lattigo}, December 2019.
\newblock EPFL-LDS.

\bibitem[MG{\etalchar{+}}11]{MN2011}
Peter Mell, Tim Grance, et~al.
\newblock The nist definition of cloud computing.
\newblock 2011.

\bibitem[Mon85]{MM1985}
Peter~L Montgomery.
\newblock Modular multiplication without trial division.
\newblock {\em Mathematics of computation}, 44(170):519--521, 1985.

\bibitem[NLV11]{NC2011}
Michael Naehrig, Kristin Lauter, and Vinod Vaikuntanathan.
\newblock Can homomorphic encryption be practical?
\newblock In {\em Proceedings of the 3rd ACM workshop on Cloud computing
  security workshop}, pages 113--124. ACM, 2011.

\bibitem[PH10]{PC2010}
Kre{\v{s}}imir Popovi{\'c} and {\v{Z}}eljko Hocenski.
\newblock Cloud computing security issues and challenges.
\newblock In {\em The 33rd International Convention MIPRO}, pages 344--349.
  IEEE, 2010.

\bibitem[PJ03]{PE2003}
Sungmee Park and Sundaresan Jayaraman.
\newblock Enhancing the quality of life through wearable technology.
\newblock {\em IEEE Engineering in medicine and biology magazine},
  22(3):41--48, 2003.

\bibitem[RAD{\etalchar{+}}78]{RD1978}
Ronald~L Rivest, Len Adleman, Michael~L Dertouzos, et~al.
\newblock On data banks and privacy homomorphisms.
\newblock {\em Foundations of secure computation}, 4(11):169--180, 1978.

\bibitem[SJJT86]{SR1986}
Michael~A Soderstrand, W~Kenneth Jenkins, Graham~A Jullien, and Fred~J Taylor.
\newblock {\em Residue number system arithmetic: modern applications in digital
  signal processing}.
\newblock IEEE press, 1986.

\bibitem[ST67]{SR1967}
Nicholas~S Szabo and Richard~I Tanaka.
\newblock {\em Residue arithmetic and its applications to computer technology}.
\newblock McGraw-Hill, 1967.

\bibitem[Tec19]{PL2019}
Duality Technologies.
\newblock \uppercase{Palisade} library.
\newblock 2019.

\bibitem[Vin16]{VE2016}
Ivan~Matveevich Vinogradov.
\newblock {\em Elements of number theory}.
\newblock Courier Dover Publications, 2016.

\end{thebibliography}

\end{document}